\documentclass{article}

\PassOptionsToPackage{numbers,compress}{natbib}
\usepackage[preprint]{neurips_2026}

\usepackage[utf8]{inputenc}
\usepackage[T1]{fontenc}
\usepackage{hyperref}
\usepackage{url}
\usepackage{booktabs}
\usepackage{graphicx}
\usepackage{amsfonts}
\usepackage{amsmath}
\usepackage{microtype}
\usepackage{xcolor}
\usepackage{tabularx}
\usepackage{enumitem}
\usepackage{subcaption}

\hypersetup{
  hidelinks,
  pdftitle={Sibyl-AutoResearch: Autonomous Research Needs Self-Evolving Trial-and-Error Harnesses, Not Paper Generators},
  pdfauthor={Chengcheng Wang, Qinhua Xie, Wei He, Jianyuan Guo, Shiqi Wang, Chang Xu}
}
\graphicspath{{figures/final/}}
\setlist[itemize]{leftmargin=*, itemsep=1pt, topsep=2pt}
\setlist[enumerate]{leftmargin=*, itemsep=1pt, topsep=2pt}
\newcolumntype{Y}{>{\raggedright\arraybackslash}X}
\newcommand{\sibyl}{\textsc{Sibyl}}

\title{Sibyl-AutoResearch: Autonomous Research Needs Self-Evolving Trial-and-Error Harnesses, Not Paper Generators}

\author{%
  \begin{tabular}{@{}c@{}}
  Chengcheng Wang\textsuperscript{1*} \quad Qinhua Xie\textsuperscript{2*} \quad Wei He\textsuperscript{3}\\
  Jianyuan Guo\textsuperscript{4} \quad Shiqi Wang\textsuperscript{4} \quad Chang Xu\textsuperscript{1}\\[0.4em]
  {\normalfont
  \textsuperscript{1}University of Sydney \quad \textsuperscript{2}East China Normal University}\\[-0.1em]
  {\normalfont
  \textsuperscript{3}TokenRhythm AI \quad \textsuperscript{4}City University of Hong Kong}\\[0.4em]
  {\normalfont
  \texttt{cwan0785@uni.sydney.edu.au}\hspace{0.5em}\texttt{10214102413@stu.ecnu.edu.cn}}\\[-0.1em]
  {\normalfont
  \texttt{wei.he@tokenrhythm.ai}\hspace{0.5em}\texttt{jianyguo@cityu.edu.hk}\hspace{0.5em}\texttt{shiqwang@cityu.edu.hk}}\\[-0.1em]
  {\normalfont
  \texttt{c.xu@sydney.edu.au}}
  \end{tabular}
}

\begin{document}

\maketitle
\begingroup
\renewcommand{\thefootnote}{\fnsymbol{footnote}}
\footnotetext[1]{Equal contribution.}
\endgroup

\begin{abstract}
Autonomous research systems increasingly make the scientific workflow executable: agents can propose ideas, run code, inspect results, and draft papers. But executable workflows do not by themselves produce research judgment. We analyze where current systems lose trial experience: weak evidence becomes prose, pilot signals become broad claims, memory remains textual, and recurring process failures do not change later behavior.
We introduce \emph{Sibyl-AutoResearch}, a self-evolving AutoResearch framework built around Scientific Trial-and-Error Harnesses. A harness lets agents run bounded trials, preserve positive and negative outcomes, and route lessons into later planning, validation, claim scope, scheduling, critique, writing, and harness repair. We formalize this through two auditable conversion units: \emph{trial-to-behavior conversion}, which links trial signals to later research actions, and \emph{trial-to-harness-behavior conversion}, which links recurring process failures to system updates.
We implement the framework in \sibyl{}, a file-backed autonomous research system that exposes the state, roles, memory, gates, and artifact traces needed to inspect these conversion paths. A retrospective audit identifies eight high-confidence conversion events, with a median latency of one iteration and a maximum latency of three iterations. A recovered-failure registry further shows how five naturally occurring failure classes, including duplicate results, stale numbers, and unsupported statistics, were blocked, downgraded, or routed into later repair. These traces do not establish a comparative performance claim; they show that the proposed conversion units are recoverable from realistic autonomous-research workspaces. The \sibyl{} framework and system are available at \href{https://github.com/Sibyl-Research-Team/AutoResearch-SibylSystem}{https://github.com/Sibyl-Research-Team/AutoResearch-SibylSystem}.
 
\end{abstract}

\section{Introduction}

Autonomous research is becoming a concrete systems problem. LLM-driven agents~\citep{wu2023autogen,lu2024aiscientistfullyautomated,boiko2023autonomouschemicalresearch} can already read papers, write code, call tools, run experiments, revise drafts, and keep long contexts. The harder question is no longer only whether these agents can assist a human researcher. It is whether an agent can improve as a researcher across attempts: form better priors, notice fragile evidence earlier, stop bad stories before they become claims, and carry hard lessons into the next project.

This is also how human research practice develops. A researcher who has spent months with a benchmark often knows which metric is brittle, which baseline result is suspicious, which negative result is useful, and which pilot result is not ready for a paper. This ability is often called intuition or taste. We call it \emph{research judgment}: experience-backed behavior that changes what the researcher does next. The point is not that judgment is mysterious, but that it is produced by many concrete trials, mistakes, repairs, and reviews.

Recent autonomous research systems automate many parts of the research loop. For example, end-to-end AI scientist frameworks proceed from idea generation to draft papers~\citep{lu2024aiscientistfullyautomated,yamada2025aiscientistv2workshoplevelautomated}, while research assistant and co-scientist systems support hypothesis formation, literature synthesis, and candidate ranking~\citep{schmidgall2025agentlaboratoryusingllm,gottweis2025aicoscientist,ghareeb2025robinmultiagentautomatingscientific}. Metric-driven loops can further search over code or methods when an explicit score is available~\citep{karpathy2026autoresearch}. Together, these systems show that agents can participate in scientific work, but they also expose a deeper design gap: \textbf{completing research stages is not the same as accumulating research judgment.} The gap becomes visible in the failures that occur after useful signals have already been observed. A pilot may expose a broken metric, but the next plan still relies on it. A reviewer objection may identify an unsupported claim, but the writer only polishes the claim. A failed GPU run may reveal a wasteful experiment order, but the scheduler repeats it. A reflection file may record the right lesson, but the planner, critic, or supervisor never receives it. \textbf{These are not just bugs; they are missing update paths from trial history to later action.}

We address this missing-update-path problem with \textbf{Sibyl-AutoResearch}, a self-evolving AutoResearch framework built around Scientific Trial-and-Error Harnesses, and we instantiate it in \sibyl{}, a file-backed autonomous research system. By harness, we mean the research environment around the agent: its state, tools, roles, memory, gates, artifact contracts, compute policies, and repair mechanisms. It lets agents try ideas under bounded controls, preserve outcomes, and convert trial history into later action. In a strong harness, past trials change future research behavior, while recurring process failures change the harness itself. These two feedback loops form agent-harness co-evolution.

The \sibyl{} system is not only a motivating example. It is the concrete implementation through which the framework was refined and audited. In \sibyl{}, research state, plans, role outputs, experiment artifacts, reviews, reflections, and writing products are stored as inspectable files. Early system runs repeatedly preserved useful signals without routing them into the next plan, claim boundary, validation gate, or scheduling policy. Those failures forced the framework to become more explicit about conversion units, role-specific routing, and harness self-evolution.

\begin{figure*}[t]
  \centering
  \begin{subfigure}[t]{0.485\textwidth}
    \centering
    \includegraphics[width=0.97\linewidth,height=0.20\textheight,keepaspectratio]{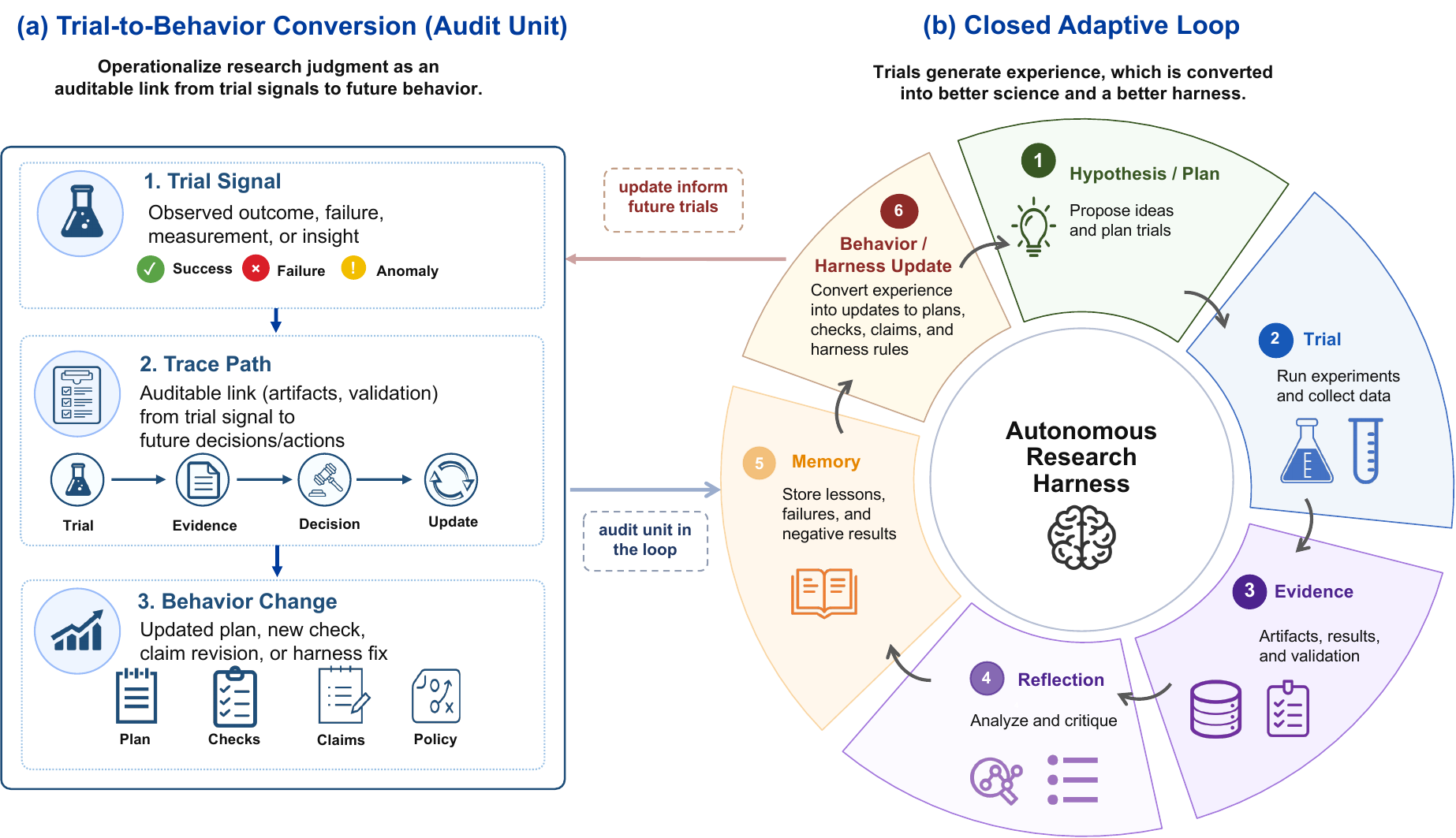}
    \caption{\small{Trial-to-behavior conversion and the closed adaptive harness loop.}}
    \label{fig:hook}
  \end{subfigure}
  \hfill
  \begin{subfigure}[t]{0.485\textwidth}
    \centering
    \includegraphics[width=0.97\linewidth,height=0.20\textheight,keepaspectratio]{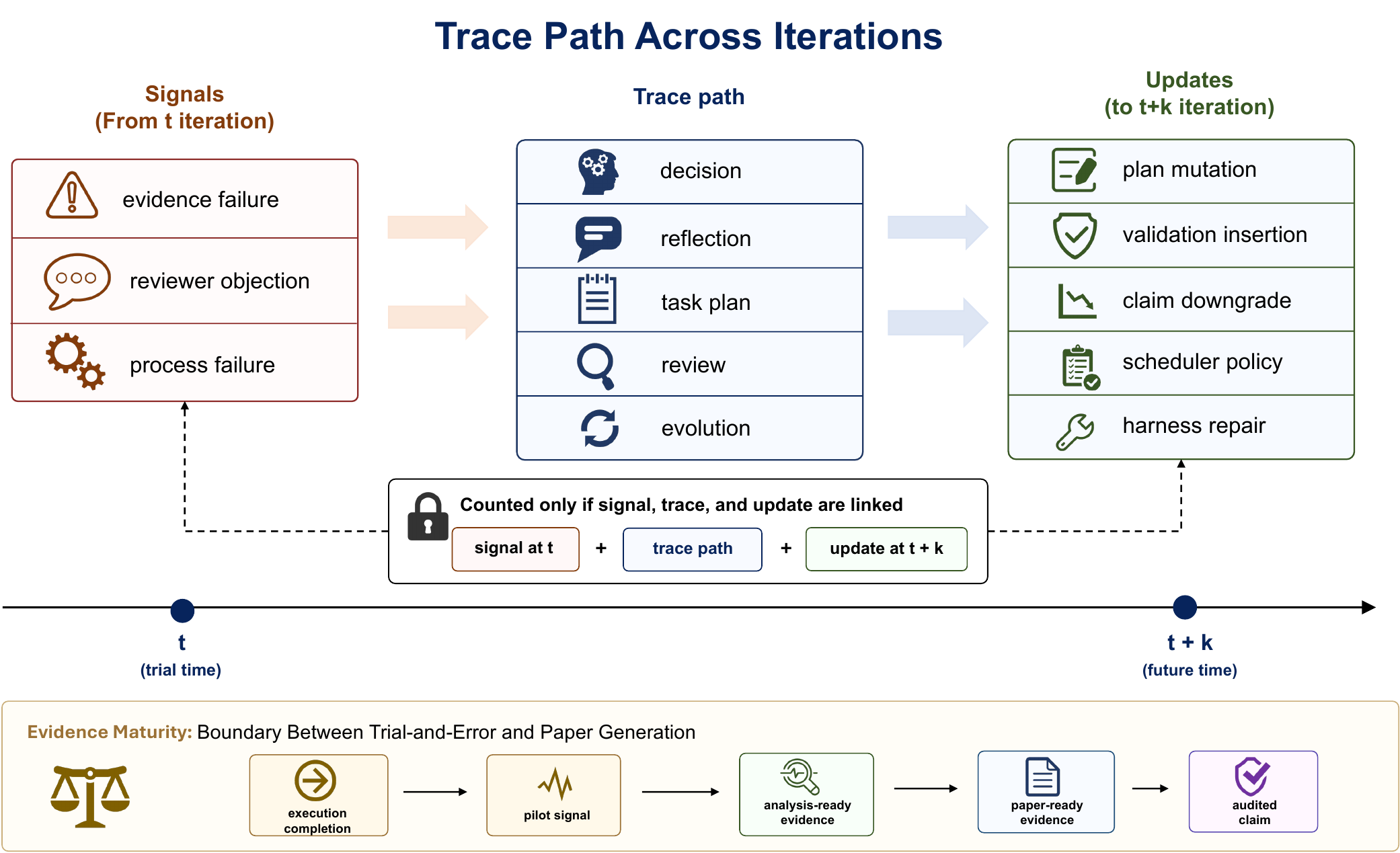}
    \caption{\small{Auditable conversion events and the evidence maturity boundary.}}
    \label{fig:conversionevent}
  \end{subfigure}
  \caption{\small{Two views of Scientific Trial-and-Error Harnesses. (a) An auditable conversion event links a trial signal, a trace path, and a later behavior change inside a closed adaptive harness loop. (b) A conversion is counted only when a signal at iteration $t$, a trace path, and an update at $t+k$ are linked; the maturity ladder separates execution completion, pilot signals, analysis-ready evidence, paper-ready evidence, and audited claims.}}
\end{figure*}

We operationalize this argument with two auditable conversion units. \emph{Trial-to-behavior conversion} asks whether a trial signal at iteration $t$ changes an action at iteration $t+k$, such as a plan, validation, claim boundary, schedule, critique, or writing scope. \emph{Trial-to-harness-behavior conversion} asks whether repeated process failures change gates, prompt overlays, telemetry requirements, scheduler policies, repair tasks, or protected constraints. The unit is small on purpose: a signal, a trace path, and a later behavior change. Figure~\ref{fig:hook} shows the audit unit inside the adaptive harness loop.

The contributions are summarized as follows:
\begin{enumerate}[leftmargin=*, itemsep=0pt, topsep=0pt]
  \item \textbf{A failure analysis for autonomous research.} We identify six recurring ways existing systems lose trial experience before it becomes later research behavior.
  \item \textbf{Two auditable conversion units.} We define trial-to-behavior and trial-to-harness-behavior conversion as inspectable links from trial signals to later behavior and harness updates.
  \item \textbf{The Sibyl-AutoResearch framework.} We distill seven harness functions and observable commitments for preserving evidence, routing memory, separating perspectives, managing compute, and repairing recurring failure paths.
  \item \textbf{A concrete \sibyl{} system.} We describe \sibyl{}, a file-backed autonomous research system that implements the framework and preserves traces for stress-testing whether the proposed conversion units are observable in realistic autonomous-research workflows.
\end{enumerate}

\section{Failure modes: where autonomous research loses experience}

Autonomous research systems already make many research actions executable. The core failure is subtler. A system can propose hypotheses, run code, optimize a metric, preserve logs, and write a draft while still losing the experience that those actions should have produced. We call the missing route an \emph{update path}: the path by which a trial signal becomes a later constraint on planning, validation, claim scope, resource allocation, critique, writing, or harness behavior.

\textbf{F1: Paper completion hides evidence immaturity.}
A pipeline can finish a paper even when the evidence is weak or corrupted. The needed update path is from weak evidence to a claim boundary: the writer should downgrade or remove the claim, and the planner should receive a validation task.

\textbf{F2: Pilot signals collapse into paper claims.}
A cheap pilot can be useful because it points to a direction. It does not by itself support a broad scientific claim. The needed update path is from pilot signal to maturity state: the system should mark the result as exploratory and require a stronger full-scale gate before writing general conclusions.

\textbf{F3: Visible objectives become bad objectives.}
Metric-driven loops work well when the objective is trusted. Research metrics often are not. A broken metric should trigger measurement critique, control design, and baseline audit. If the system simply optimizes the same metric again, the trial did not become research judgment.

\textbf{F4: Memory stays textual instead of routed.}
Long context can preserve a lesson in text while failing to route it to the role that needs it. The needed update path is from lesson to role-specific behavior: planner checks, critic objections, supervisor gates, scheduler policy, or writer restrictions.

\textbf{F5: More trials do not improve trial policy.}
Many-trial systems can increase experiment volume without improving the order, cost, stopping rules, or sanity checks. A failed or wasteful run should change resource allocation, early-stop policy, or cheap-check ordering.

\textbf{F6: Process failures recur because the harness does not change.}
Missing artifacts, stale tables, incomplete telemetry, and paper/evidence synchronization errors are not only project-local mistakes. Repeated process failures should change the harness: new gates, prompt overlays, artifact contracts, telemetry requirements, repair tasks, or protected constraints.

These failure modes are distilled from recurring patterns in \sibyl{} workspace traces and from comparison with prior autonomous-research designs. They are not presented as exhaustive. Their role is to identify where paper-completion systems lose experience and where a harness must expose an update path. For example, in one sparse-autoencoder replication run, a strong writing score coexisted with duplicate result files, a feature-count mismatch, a missing sparsity-matched control, and many inactive sparse features. In an image-augmentation pilot, a CIFAR-10 direction looked promising, but full-scale claims were blocked because the larger follow-up experiments were still missing. In diffusion-language-model acceleration and dynamic weight-decay projects, unsupported statistics, metric failures, and control problems became later claim downgrades, validation tasks, and harness repairs rather than polished prose.

Table~\ref{tab:failuremodes} restates the six failure modes as missing update paths. Each row points to a positive harness function in Section~\ref{sec:harness}.

\begin{table}[t]
  \centering
  \footnotesize
  \setlength{\tabcolsep}{2.5pt}
  \caption{Common autoresearch failure modes restated as missing update paths.}
  \label{tab:failuremodes}
  \resizebox{\linewidth}{!}{%
  \begin{tabularx}{\linewidth}{p{0.20\linewidth}p{0.30\linewidth}Y}
    \toprule
    Failure mode & Missing update path & Why it matters \\
    \midrule
    F1: Paper completion & Evidence signals do not constrain claims & Polished writing can absorb uncertainty without sending weak claims back to validation. \\
    F2: Pilot collapse & Pilot signals do not change evidence maturity & A direction worth testing is mistaken for a result worth stating. \\
    F3: Bad objective & Measurement failures do not revise the objective & Optimizing an easy proxy can move the system away from understanding. \\
    F4: Unrouted memory & Lessons do not reach the role that needs them & The system can restate a failure without changing the next plan, critique, or gate. \\
    F5: Trial-volume drift & Negative or wasteful trials do not revise resource policy & Trial volume increases without better stopping rules, ordering, or scope. \\
    F6: Static harness & Process failures do not repair the harness & The same artifact, telemetry, or synchronization failure can recur across projects. \\
    \bottomrule
  \end{tabularx}
  }
\end{table}

\section{Existing systems and remaining gaps}

\textbf{End-to-end autonomous research systems.}
AI scientist systems show that LLM agents can generate ideas, run experiments, interpret results, and draft manuscripts \citep{lu2024aiscientistfullyautomated,yamada2025aiscientistv2workshoplevelautomated}. Agent laboratory and co-scientist systems emphasize human collaboration, literature synthesis, hypothesis generation, and candidate ranking \citep{schmidgall2025agentlaboratoryusingllm,gottweis2025aicoscientist,ghareeb2025robinmultiagentautomatingscientific}. Domain-specific scientific agents further show that language-model systems can combine literature search, code execution, tool use, and laboratory or chemistry-specific automation \citep{bran2023chemcrow,boiko2023autonomouschemicalresearch}. More recent systems extend this trajectory toward longer-horizon and more domain-grounded discovery: cmbagent uses a planning-and-control multi-agent architecture for an autonomous cosmology analysis task, Kosmos coordinates data analysis and literature search through a structured world model, SAGA evolves scientific objective functions rather than treating objectives as fixed, and Aster accelerates iterative program-improvement loops across scientific and engineering tasks \citep{xu2025opensourceplanning,mitchener2025kosmos,du2025saga,bicker2026aster}. MLAgentBench makes the experimentation loop itself an evaluation target by asking agents to improve machine-learning systems through file edits, execution, and result inspection \citep{huang2023mlagentbench}. General-purpose multi-agent and software-agent systems also show how roles, messages, tools, human interaction, and agent-computer interfaces shape task execution \citep{wu2023autogen,hong2023metagpt,yang2024sweagent,jimenez2023swebench}. These systems make more of the scientific workflow executable, but the failure modes above show why execution alone is not enough: trial signals must change later research behavior and the harness that hosts later trials.

\textbf{Metric-driven search and verifier-rich discovery.}
Autoresearch loops, FARS-style systems, AlphaEvolve-style systems, Aster-style program improvement, CORAL-style multi-agent evolution, and PaperBench-like evaluations show the value of repeated trials when feedback is clear \citep{karpathy2026autoresearch,analemma2026fars,novikov2025alphaevolvecodingagentscientific,bicker2026aster,qu2026coral,starace2025paperbenchevaluatingaisability}. This line of work is closest to classical AutoML, neural architecture search, and black-box optimization: the system searches over candidates under an explicit objective \citep{hutter2019automated,zoph2017neuralarchitecturesearchreinforcement}. The difference is that open-ended research often lacks a single trusted objective. A harness must therefore route measurement failures and negative results into revised metrics, controls, and claim boundaries, not only into another search step.

\textbf{Agent memory, reflection, and harness infrastructure.}
Tool-using agents establish the basic pattern of interleaving language reasoning with external actions, while reflection and memory systems show that lessons from failed trials can improve later behavior \citep{yao2022react,schick2023toolformer,shinn2023reflexionlanguageagentsverbal,packer2024memgptllmsoperatingsystems,madaan2023selfrefineiterativerefinementselffeedback,wang2023voyageropenendedembodiedagent,park2023generativeagentsinteractivesimulacra}. Broad agent benchmarks and software or research benchmarks expose the importance of execution, long-context state, and artifact inspection \citep{liu2023agentbench,jimenez2023swebench,huang2023mlagentbench,starace2025paperbenchevaluatingaisability}. Long-running agents require harnesses, tools, tracing, and guardrails to make progress inspectable and safe \citep{anthropic2025harnesses,anthropic2025tools,openai2026tracing,openai2026guardrails}. Scientific provenance and reproducibility work gives complementary tools for connecting claims to artifacts \citep{733f89c65e4844f9aabcae1c276a5602,Gundersen_2018}. Work on weak evidence and publication incentives warns that polished claims can outrun the evidence base \citep{Ioannidis_2005}. We connect these threads by treating a harness as a memory-bearing research environment with explicit evidence boundaries and by requiring claim-relevant behavior changes to be visible in traces.

\textbf{Expertise and research judgment.}
The idea that expertise grows from repeated, feedback-rich experience is old \citep{Ericsson_1996}. Research judgment in this paper is the system-level analogue of that process. We do not claim that agents acquire human expertise in the psychological sense. We make a narrower systems claim: if an autonomous research system has learned from a trial, later behavior should expose that learning.

\section{The Sibyl-AutoResearch framework}
\label{sec:harness}

The diagnosis above changes the unit of analysis and motivates the Sibyl-AutoResearch framework. The central unit of autonomous research is the \emph{trial}: a bounded encounter with a real research environment that produces a signal about a hypothesis, method, measurement, baseline, validation check, resource policy, or process. A trial is valuable when the signal changes later behavior. A failed trial can be especially useful because it rules out a tempting story, exposes a fragile metric, or reveals a weakness in the harness.

Sibyl-AutoResearch treats a Scientific Trial-and-Error Harness as the environment that makes those updates possible. It is a set of harness functions around the agent: state, tools, roles, memory, gates, artifact contracts, compute control, and repair mechanisms. These functions do not guarantee good science. They make research behavior inspectable at the places where judgment should appear.

Although the framework is presented abstractly in this section, it was refined through system-building pressure from \sibyl{}. Whenever a \sibyl{} run preserved a useful signal without changing later behavior, we treated the failure as evidence that the framework needed a more explicit update path. The seven functions below are therefore design commitments for AutoResearch systems, not claims that any current harness fully solves autonomous research.

\textbf{H1: Trial orchestration.}
Each trial should have a question, expected evidence, dependencies, outputs, and stop conditions. The observable commitment is that earlier evidence changes the next trial plan, branch priority, or task dependency.

\textbf{H2: Evidence maturity.}
Execution completion, pilot signal, analysis-ready evidence, paper-ready evidence, and audited claim are different states. The observable commitment is that a claim advances only after validation, scope control, and artifact links. Negative evidence can increase maturity by ruling out a false story.

\textbf{H3: Traceability.}
A behavior update should be tied to artifacts: plans, configs, logs, tables, reviews, negative evidence, and writing changes. The observable commitment is that a reader can reconstruct why a later action changed.

\textbf{H4: Routed research memory.}
Reflection should not remain as a free-form note. Lessons must be routed to planners, experimenters, critics, supervisors, schedulers, or writers. The observable commitment is that a past lesson changes a later role-specific check or decision.

\textbf{H5: Perspective separation.}
Optimistic, skeptical, methodological, supervisory, and writing roles should have different authority. The observable commitment is that objections become validation tasks, plan mutations, stopped branches, or claim downgrades rather than free-form disagreement.

\textbf{H6: Resource-aware trial policy.}
Research trial-and-error is bounded by GPU time, token budget, and human review attention. The observable commitment is that failed or wasteful trials change sanity-check policy, allocation, monitoring, or recovery behavior.

\textbf{H7: Harness self-evolution with protected constraints.}
Some trial signals are about the research question. Others are about the harness that produced the research. Repeated missing artifacts, stale outputs, telemetry gaps, or gate failures should change prompt overlays, validation gates, artifact contracts, scheduler policies, or repair tasks. The observable commitment is that self-evolution strengthens evidence integrity rather than optimizing polish, pass rate, or reviewer gaming.

\textbf{Trial-to-behavior conversion} is the agent-side audit unit. A signal at iteration $t$ must alter an action at iteration $t+k$: which direction to try, which evidence to distrust, which validation to run earlier, how strongly to state a claim, when to stop a branch, how to allocate GPU budget, or when to delay writing. A conversion links three artifacts: a trial signal, a trace path, and a later behavior change. Figure~\ref{fig:conversionevent} shows this unit.

\textbf{Trial-to-harness-behavior conversion} is the harness-side audit unit. A recurring process failure must alter a harness function: a gate, prompt overlay, telemetry requirement, scheduler policy, repair task, artifact contract, or protected constraint. These two conversion units define agent-harness co-evolution: the agent accumulates research judgment, and the harness learns how to make later research loops safer, cheaper, and more informative.

Table~\ref{tab:evaluation} maps the six failure modes to the seven positive commitments above.

\begin{table}[t]
  \centering
  \footnotesize
  \setlength{\tabcolsep}{2.5pt}
  \caption{Observable commitments in the Sibyl-AutoResearch framework.}
  \label{tab:evaluation}
  \resizebox{\linewidth}{!}{%
  \begin{tabularx}{\linewidth}{p{0.23\linewidth}p{0.20\linewidth}p{0.32\linewidth}Y}
    \toprule
    Harness function & Primary failure mode addressed & What must be observable & Failure it rules out \\
    \midrule
    H1 Trial orchestration & F5 Trial-volume drift & Earlier evidence changes later plans, dependencies, branch priority, or stopping decisions. & More activity without better trial policy. \\
    H2 Evidence maturity & F1 Paper completion; F2 Pilot collapse & Claims move through explicit maturity states with validation and scope labels. & Completed workflows that overclaim weak, stale, or corrupted evidence. \\
    H3 Traceability & F1 Paper completion & Claims and behavior updates link to scripts, configs, logs, results, reviews, and negative evidence. & Well-written claims whose evidence path cannot be reconstructed. \\
    H4 Routed memory & F4 Unrouted memory & Lessons change checks or priors of the roles that need them. & Long context that remembers text but does not change action. \\
    H5 Perspective separation & F3 Bad objective; F4 Unrouted memory & Objections become validation tasks, plan mutations, stopped branches, or claim downgrades. & Critique that remains theater. \\
    H6 Resource policy & F5 Trial-volume drift & Failed or wasteful trials change sanity-check order, allocation, monitoring, or recovery. & Trial volume that wastes compute and review budget. \\
    H7 Harness self-evolution & F6 Static harness & Recurring process failures change gates, overlays, telemetry, scheduler policy, repair tasks, or protected constraints. & The same infrastructure failure recurring across projects. \\
    \bottomrule
  \end{tabularx}}
\end{table}

\section{The \sibyl{} system}
\label{sec:sibyl}

\sibyl{} is the concrete system realization of Sibyl-AutoResearch. It is a file-backed autonomous research system in which research state, plans, roles, memory, gates, experiment artifacts, reviews, and writing outputs are preserved as inspectable files rather than hidden runtime state. Early versions of the system exposed the same failure pattern repeatedly: useful signals were preserved as files but did not reliably change planning, validation, claim scope, scheduling, critique, or writing authority. The framework made those failures nameable, and later \sibyl{} mechanisms were instrumented to preserve the traces needed to audit them.

We do not present \sibyl{} as a controlled benchmark against prior systems or as comparative performance evidence. Its role in this paper is both architectural and methodological: it shows how the proposed AutoResearch framework can be implemented in a real autonomous-research environment, and it provides the trace substrate used to audit the proposed conversion units.

The current implementation reflects the seven harness functions in Section~\ref{sec:harness}. Trial orchestration uses an artifact-backed state machine and task plans. Evidence maturity is enforced through decisions, quality gates, validation requirements, and writing restrictions. Traceability comes from workspace artifacts, event logs, reviews, experiment state, and writing outputs. Routed memory uses reflection postprocessing, evolution records, issue categories, and role-specific lesson overlays. Separate planner, experimenter, critic, supervisor, skeptic, methodologist, writer, and editor roles keep objections from being silently absorbed into prose. GPU scheduling, dependency layers, monitoring, recovery, repair tasks, self-heal mechanisms, and protected constraints provide resource policy and harness self-evolution.

Two implementation boundaries are especially important. First, reflection outputs are normalized into issue categories, converted into evolution records, and injected as role-specific lesson overlays rather than left as long-context text. Second, writing agents consume a validated claim registry with maturity labels, artifact links, and validation status; pilot signals remain usable as pilot signals but cannot be upgraded into paper-ready claims by prose alone.

Table~\ref{tab:sibylmapping} maps each framework commitment to the current \sibyl{} mechanism that makes the corresponding update path inspectable. The mapping is descriptive, not a claim of completeness. Appendix~\ref{sec:appendix-implementation} expands these mechanisms and supporting diagrams. Section~\ref{sec:process-evidence} then asks whether the preserved traces contain actual conversion events.

\begin{table}[t]
  \centering
  \footnotesize
  \setlength{\tabcolsep}{2.5pt}
  \caption{How the current \sibyl{} system operationalizes the Sibyl-AutoResearch commitments.}
  \label{tab:sibylmapping}
  \begin{tabularx}{\linewidth}{p{0.25\linewidth}p{0.31\linewidth}Y}
    \toprule
    Harness function & \sibyl{} mechanism & Observable update enabled \\
    \midrule
    H1 Trial orchestration & State machine and task plans & Earlier evidence changes plans, dependencies, pilot/full transitions, or branch priority. \\
    H2 Evidence maturity & Decision gates and validation requirements & Weak evidence triggers a refinement decision, pivot, claim downgrade, or writing restriction. \\
    H3 Traceability & Workspace artifacts and event logs & Claims and updates trace to plans, logs, results, reviews, and negative evidence. \\
    H4 Routed memory & Reflection, evolution records, and role overlays & Recurring issues become checks for planners, experimenters, critics, supervisors, or writers. \\
    H5 Perspective separation & Role-separated agents and debates & Objections become validation tasks, plan mutations, or scoped claims. \\
    H6 Resource policy & GPU scheduler, dependency graph, and recovery & Failed or wasteful trials change sanity-check policy, allocation, or monitoring. \\
    H7 Harness self-evolution & Evolution records, self-heal tasks, and protected constraints & Process failures become gates, telemetry requirements, overlays, or repair tasks. \\
    \bottomrule
  \end{tabularx}
\end{table}

This implementation detail matters because writing and research roles do not have the same authority. A writer can synthesize validated claims but cannot upgrade a pilot result. A supervisor can allow advancement only with scoped risks. A critic's objection must become a task or boundary condition. A scheduler can run expensive experiments, but low evidence value should trigger cheap checks first. These boundaries are scientific integrity mechanisms, not only engineering choices.

\section{Evidence from the \sibyl{} system}
\label{sec:process-evidence}

We use preserved \sibyl{} workspaces to ask a narrower question than system performance: can the proposed conversion units be recovered from realistic autonomous-research traces? This is a retrospective process audit, not a controlled comparison and not an estimate of average harness effectiveness. We mark 8 high-confidence conversion events across the workspace set and estimate a median latency of 1 iteration, with a maximum visible latency of 3 iterations, from signal to behavior update. The count is hand-audited and conservative. It is useful only as evidence of inspectability; it should not be read as a benchmark score or a claim that \sibyl{} converts every useful signal.

For readability, the paper names cases by the research problem or failure type rather than by internal workspace directories. Appendix~\ref{sec:extended-cases} gives additional case summaries and artifact categories without exposing internal file paths.

The evidence is organized in three layers. First, hand-audited conversion events test the mechanism directly: did a signal change a later plan, validation task, claim boundary, schedule, critique, or writing restriction? Table~\ref{tab:conversionevents} summarizes this conservative event sample. Second, the recovered-failure registry asks whether naturally occurring evidence-boundary failures were blocked, downgraded, or routed into repair. Third, an aggregate review-to-action audit asks whether reviewer-like objections become later experiments, validation, or harness changes rather than another score to optimize. The paper's claim is not that \sibyl{} drafts receive high review scores; it is that objections and failures convert into later research or harness behavior.

\begin{table}[t]
  \centering
  \small
  \setlength{\tabcolsep}{2.7pt}
  \caption{Audited conversion-event sample. Latency is measured in iterations when a later update iteration is visible.}
  \label{tab:conversionevents}
  \resizebox{\linewidth}{!}{%
  \begin{tabularx}{\linewidth}{p{0.25\linewidth}p{0.29\linewidth}p{0.24\linewidth}>{\centering\arraybackslash}p{0.07\linewidth}Y}
    \toprule
    Case & Trial signal & Later behavior update & Latency & Harness function \\
    \midrule
    Dynamic weight-decay control & Controller instability, compute-budget confounds, and hidden negative controls. & Repaired controller, epoch-budget assertions, 9/9 stability tests, and scoped advancement. & 1 & H1, H2, H3 \\
    Diffusion-language-model acceleration & Unsupported statistics, accept-rate mismatch, and one proposed accelerator behaving as a functional no-op. & Speedup story reframed as an interference taxonomy; full-scale replication prerequisites added. & 1 & H2, H5 \\
    Sparse-autoencoder absorption & Writing stagnation and repeated missing source-to-paper validation. & Experiment-first planning and validation-first gate in later iterations. & 3 & H1, H4 \\
    Sparse-autoencoder absorption & Confidence-interval inversion, stale headline ratios, and incompatible first-letter rates. & Source-to-paper validation script, corrected aggregation, and headline ratio reduced from 4.1\(\times\) to 2.7\(\times\). & 1 & H2, H3 \\
    Failed sparse-autoencoder replication & Writing score rose while duplicate replicates, feature-count mismatch, and inactive features broke the evidence base. & Duplicate detection, feature-count verification, sparsity-matched controls, and single-source analysis became prerequisites. & 1 & H2, H3 \\
    Image-augmentation pilot & Pilot produced a go signal but full-scale evidence was absent. & Claims blocked at the pilot/full boundary and larger follow-up experiments required. & 0 & H2 \\
    Diffusion-language-model caching pilot & Caching pilot spent 54 minutes to find 15.2\(\times\) overhead. & Later policy recorded a 10-minute throughput sanity check to catch similar failures earlier. & 0 & H6, H7 \\
    Missing-review gate stress test & Review score missing at quality gate. & Iteration hard-blocked and rolled back to review. & 0 & H2, H7 \\
    \bottomrule
  \end{tabularx}}
\end{table}

\begin{figure}[t]
  \centering
  % \vspace{-2em}
  \includegraphics[width=0.95\linewidth,height=0.24\textheight,keepaspectratio]{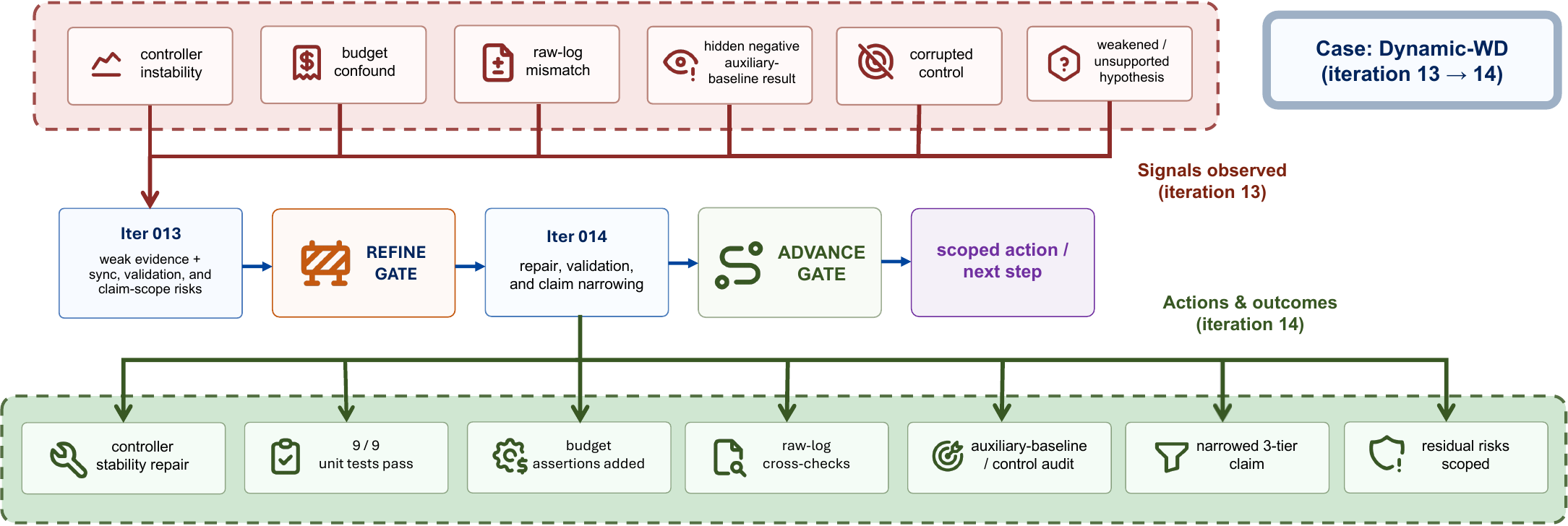}
  \caption{Dynamic weight-decay gate-to-action flow. Controller instability, budget confounds, raw-log mismatches, hidden negative auxiliary-baseline results, corrupted controls, and a weakened or unsupported hypothesis at iteration 13 trigger a refinement decision. Iteration 14 then performs controller stability repair, 9-of-9 unit tests, budget assertions, raw-log cross-checks, auxiliary-baseline/control audit, and a narrowed three-tier claim before issuing a scoped advancement decision.}
  \label{fig:dynamicwdflow}
  % \vspace{-1em}
\end{figure}

\textbf{Patterns across the audited cases.}
The detailed traces behind these conversion events are preserved in Appendix~\ref{sec:extended-cases}. Across cases, the same pattern repeats and explains why the framework emphasizes conversion rather than memory volume or paper quality: a weak, stale, or corrupted signal matters only when it becomes a later change in algorithm design, validation, claim boundary, writing permission, or harness policy. Figure~\ref{fig:dynamicwdflow} shows one concrete path. In the dynamic weight-decay case, paper errors, budget concerns, corrupted controls, and hidden negative results for an auxiliary baseline trigger a refinement decision; the next iteration adds controller repair, budget assertions, 9-of-9 stability tests, raw-log checks, and a scoped advancement decision.

\textbf{Recovered-failure registry.}
The strongest controlled experiment would inject known failures into a held-out workspace. We do not report such a new run here. Instead, we report a recovered-failure registry from naturally occurring traces. This is weaker than an injected benchmark but stronger than anecdote because each row contains a concrete failure class, an audit artifact type, a catch mechanism, and a later update. Table~\ref{tab:appendix-recoveredfailures} gives the failure rows without exposing internal file paths.

\textbf{Aggregate review-to-action audit.}
We also audited a generated-review archive containing reviewer-like artifacts over 51 project-iteration snapshots from 11 workspaces, with three review surfaces per snapshot. We include this audit as a stress test for the framework's premise, not as a validation of generated review scores. These are pressure tests, not peer-review outcomes. The most important lesson is negative: review scores are poor progress metrics. The three surfaces disagree in schema and calibration, and within each surface scores barely move across iterations even when objections persist. The objections themselves are more useful. They cluster around validation strength, claim scope, baselines, controls, reproducibility, and artifact synchronization, which are the same evidence-boundary risks the workspace traces already flag. Detailed counts and per-surface calibration are in Appendix~\ref{sec:appendix-review-transition-stats}.

\begin{figure}[t]
  \centering
  \includegraphics[width=0.98\linewidth]{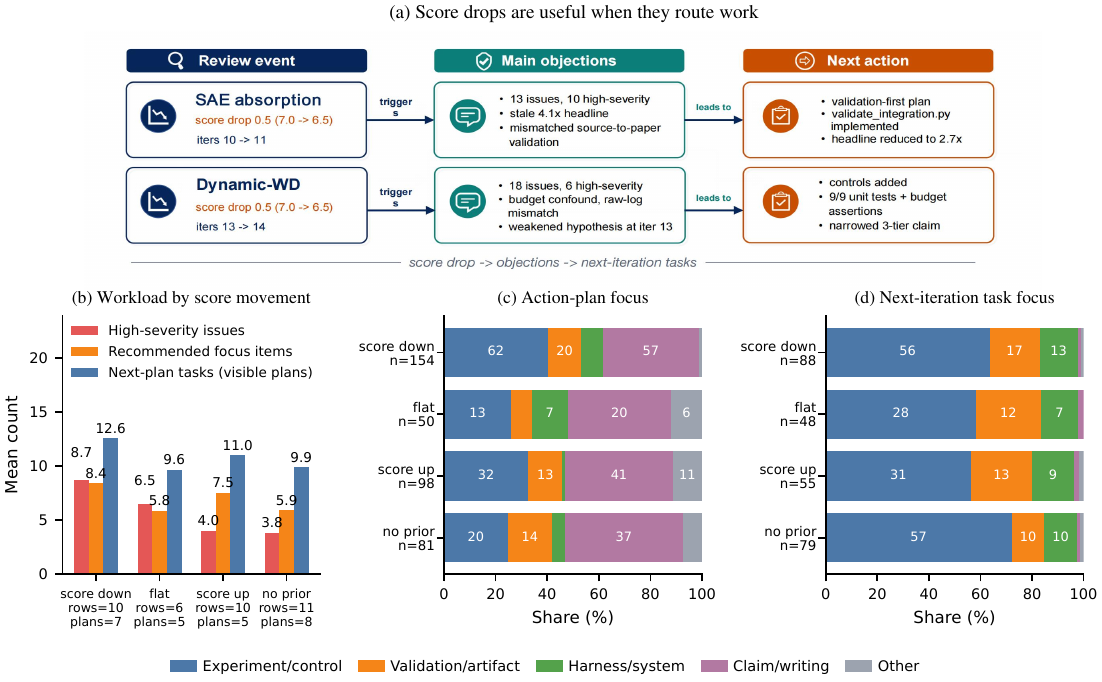}
  \caption{\small{Internal review scores as issue-to-action signals. (A) Two concrete score-drop rows show that a lower score is useful when it routes work: a sparse-autoencoder absorption drop becomes a validation-first plan and source-to-paper validation script; a dynamic weight-decay drop becomes added controls, 9/9 unit tests, and a narrowed claim. (B) Mean issue and focus load per parsed row; next-plan tasks are averaged only over rows with a visible next-iteration task plan, and the x-axis reports both parsed rows and visible-plan rows. (C--D) Heuristic multi-label classifications of structured action-plan recommendations and next-iteration task-plan entries. The post-hoc generated reviews are not causal inputs to these loops; they are pressure tests for whether similar objections would be actionable.}}
  \label{fig:scorebehaviordiag}
\end{figure}

To check whether such objections become later research items, we align internal \sibyl{} reviews, reflections, and next-iteration plans (the post-hoc external reviews are not used as inputs here). In this hand-parsed diagnostic sample, across 12 audited traces and 37 parseable structured action-plan rows, score-drop rows carry roughly twice as many high-severity issues as score-up rows (8.7 vs.\ 4.0 per row) and a heavier corrective load. When a score-drop row has a visible next plan, the next iteration is dominated by experiments and controls (about two thirds of tasks), with the rest split between validation/artifact repair and harness changes. Figure~\ref{fig:scorebehaviordiag} summarizes this review-to-action path; raw transition counts and supervisor-score calibration are in Appendix~\ref{sec:appendix-review-transition-stats}.

% \begin{figure}[t]
%   \centering
%   \vspace{-2em}
%   \includegraphics[width=0.85\linewidth]{fig_score_behavior_diagnostics.pdf}
%   \caption{Internal review scores as issue-to-action signals. (A) Two concrete score-drop rows show that a lower score is useful when it routes work: a sparse-autoencoder absorption drop becomes a validation-first plan and source-to-paper validation script; a dynamic weight-decay drop becomes added controls, 9/9 unit tests, and a narrowed claim. (B) Mean issue and focus load per parsed row; next-plan tasks are averaged only over rows with a visible next-iteration task plan, and the x-axis reports both parsed rows and visible-plan rows. (C--D) Heuristic multi-label classifications of structured action-plan recommendations and next-iteration task-plan entries. The post-hoc generated reviews are not causal inputs to these loops; they are pressure tests for whether similar objections would be actionable.}
%   \label{fig:scorebehaviordiag}
%   \vspace{-1.7em}
% \end{figure}

\textbf{Harness-side evidence.}
The \sibyl{} evolution-memory records show trial-to-harness-behavior conversion. The central digest contains 416 recurring issue patterns: 212 experiment, 89 writing, 84 analysis, 20 system, 4 ideation, 3 pipeline, 3 planning, and 1 efficiency. Routing is explicit: experiment issues are assigned to experiment-running and planning roles, analysis issues to supervisory and critique roles, and writing issues to drafting and editing roles. The prompt loader then injects selected lessons as role-specific overlays. The diffusion-language-model caching case also exposes a resource-policy update: the reflection records that a 54-minute pilot revealed a 15.2\(\times\) overhead failure that a 10-minute throughput sanity check could have caught earlier.

\section{Limitations and alternative views}
\label{sec:limitations-alternative-views}

The evidence is retrospective and concentrated in one author-built harness for computational AI/ML workspaces. The framework and \sibyl{} co-developed, so the traces are not an independent validation set for the theory. They are better understood as theory-building and stress-test evidence. The conversion count is hand-marked, the recovered-failure registry uses natural failures rather than a new injected benchmark, and the ablations are process diagnostics rather than repeated statistical experiments. The reviewer-like artifacts are generated reviews, not human peer-review decisions, and should be used only as process signals about possible evidence-boundary failures. The transition counts are raw event-log counts and may include resume or checkpoint repetition, so they support process-shape claims rather than exact execution counts. The paper supports a design framework and an existence proof, not a comparative performance claim about \sibyl{}. Future work should evaluate the conversion units on held-out harnesses, with independent annotators, prospective injected failures, and public artifact bundles.

One alternative view is that final manuscript quality is the only outcome that matters. We disagree because a manuscript is an expression of an evidence state: in the failed sparse-autoencoder replication, writing quality improved while the evidence base collapsed. The generated-review audit makes the same point at scale: reviewer-like tools disagree enough in schema and calibration that optimizing their scores would create another brittle objective.

A second view is that better verifiers and benchmarks will solve the problem. We agree where objectives are trusted, but open-ended research often discovers that the objective itself is broken; in the diffusion-language-model acceleration case, unsupported statistics and a no-op accelerator forced a change in the objective.

A third view is that human researchers should supply judgment, while agents need only provenance. Human responsibility remains central, but scalable oversight still needs traces showing why an autonomous system changed its claims and plans. A final concern is that gates and routed memory may make systems optimize process compliance. This is why the commitments emphasize protected evidence integrity, negative results, and hidden injected failures rather than a single process score.

\section{Conclusion}
Autonomous research should not be judged mainly by whether a system can generate a complete paper. A paper is only an expression of an evidence state; it does not by itself show whether the system has learned from the trials that produced it. The harder capability is whether a system can turn trial history into research judgment: better plans, stronger validation, narrower claims, safer resource policies, and a harness that becomes harder to fool over time.
This paper argues that such judgment requires Sibyl-AutoResearch: a self-evolving AutoResearch framework built around Scientific Trial-and-Error Harnesses.
Our experience building \sibyl{} suggests that the framework and the system cannot be cleanly separated. Better traces reveal missing update paths, and better update-path theory tells later system versions what to preserve, route, block, and audit. Scientific Trial-and-Error Harnesses make this agenda auditable: they ask not only what an agent produced, but what it learned, where that lesson traveled, and how it changed the next research action.

\bibliographystyle{plainnat}
\bibliography{references}

\appendix
\clearpage

\section*{Appendix}

\noindent The appendix follows the same order as the main argument. Appendix~\ref{sec:appendix-implementation} expands the \sibyl{} mechanisms and supporting diagrams. Appendix~\ref{sec:extended-cases} gives the workspace traces and the recovered-failure registry behind the conversion-event audit. Appendix~\ref{sec:appendix-review-transition-stats} reports the generated-review and event-log diagnostics. Appendix~\ref{sec:appendix-systemcomparison} gives the system-comparison details, and Appendices~\ref{sec:appendix-eval-protocols} and~\ref{sec:appendix-governance-details} collect evaluation protocols and governance details.

\vspace{0.4em}

\section{Sibyl implementation and harness diagrams}
\label{sec:appendix-implementation}

This appendix expands the implementation sketch from the main text. It is implementation evidence for how the co-developed framework was operationalized, not a full software manual or an independent proof that one harness solves autonomous research. The relevant unit is the research behavior update or harness behavior update that the mechanism makes possible.

\subsection{Reflection and evolution memory}

The memory layer starts from reflection artifacts, but it should not end there. Reflection outputs are normalized into issue categories such as system, experiment, writing, analysis, planning, pipeline, ideation, and efficiency. The evolution layer maintains digests and lessons, computes relevance, and injects selected lessons into later prompts. This is the mechanism behind the paper's distinction between long context and research memory. Long context can expose past text; routed memory can change which checks a role performs.

Memory effectiveness is measured by behavior, not volume. A lesson informs the system when it reduces repeated failures, changes a plan, adds a validation step, changes a claim boundary, or repairs the harness path that allowed the failure. The memory layer therefore stores source links, uncertainty, decay, and re-opening criteria alongside the lesson text.

\subsection{Decision gates, self-heal, and writing boundaries}

The current system exposes idea validation, experiment decisions, quality gates, review stages, and a structured self-heal substrate. The self-heal layer includes error collection, routing, protected-file constraints, deterministic fixers for recurring failure classes, and repair-task generation. These mechanisms improve workflow reliability and create places where failures become tasks before they recur in the next research loop.

The natural boundary mechanism is a validated claim registry. Writing agents consume claims with maturity labels, artifact links, and validation status. Pilot signals remain usable as pilot signals. A paper is generated from the claim registry, not from whatever narrative is most polished after the latest experiment.

\subsection{Additional mechanism diagrams}

\begin{figure}[t]
  \centering
    \vspace{6pt}
  \includegraphics[width=\linewidth,height=0.30\textheight,keepaspectratio]{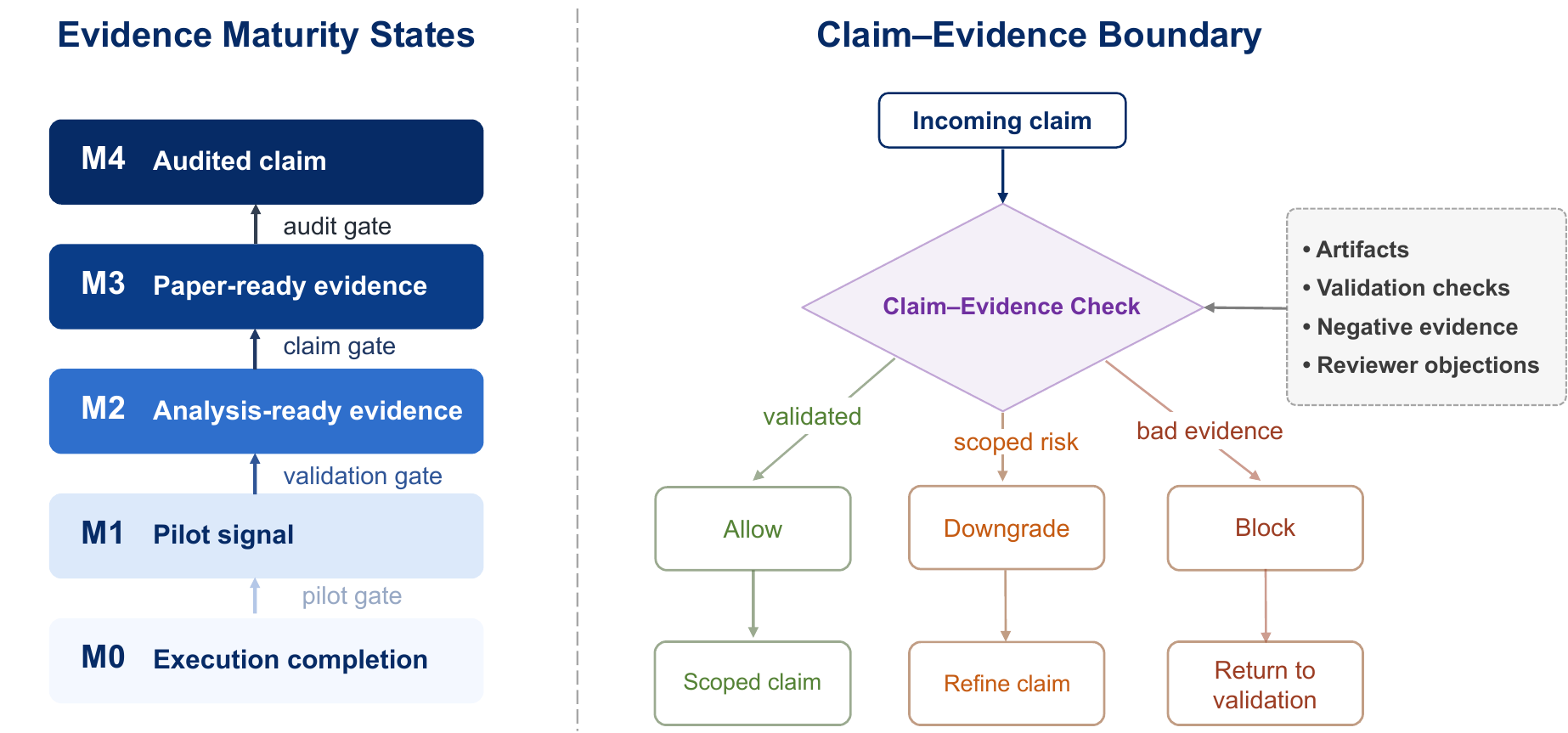}
  \caption{Evidence maturity states and the claim-evidence boundary. Execution completion, pilot signal, analysis-ready evidence, paper-ready evidence, and audited claims are separate states. The claim-evidence check allows, downgrades, or blocks claims based on artifacts, validation checks, negative evidence, and reviewer objections.}
  \label{fig:appendix-evidence-maturity}
\end{figure}

Figure~\ref{fig:appendix-evidence-maturity} expands H2 from Section~\ref{sec:harness} by showing the evidence maturity states and the claim-evidence check used to allow, downgrade, or block claims.

\begin{figure}[t]
  \centering
    \vspace{6pt}
  \includegraphics[width=\linewidth,height=0.30\textheight,keepaspectratio]{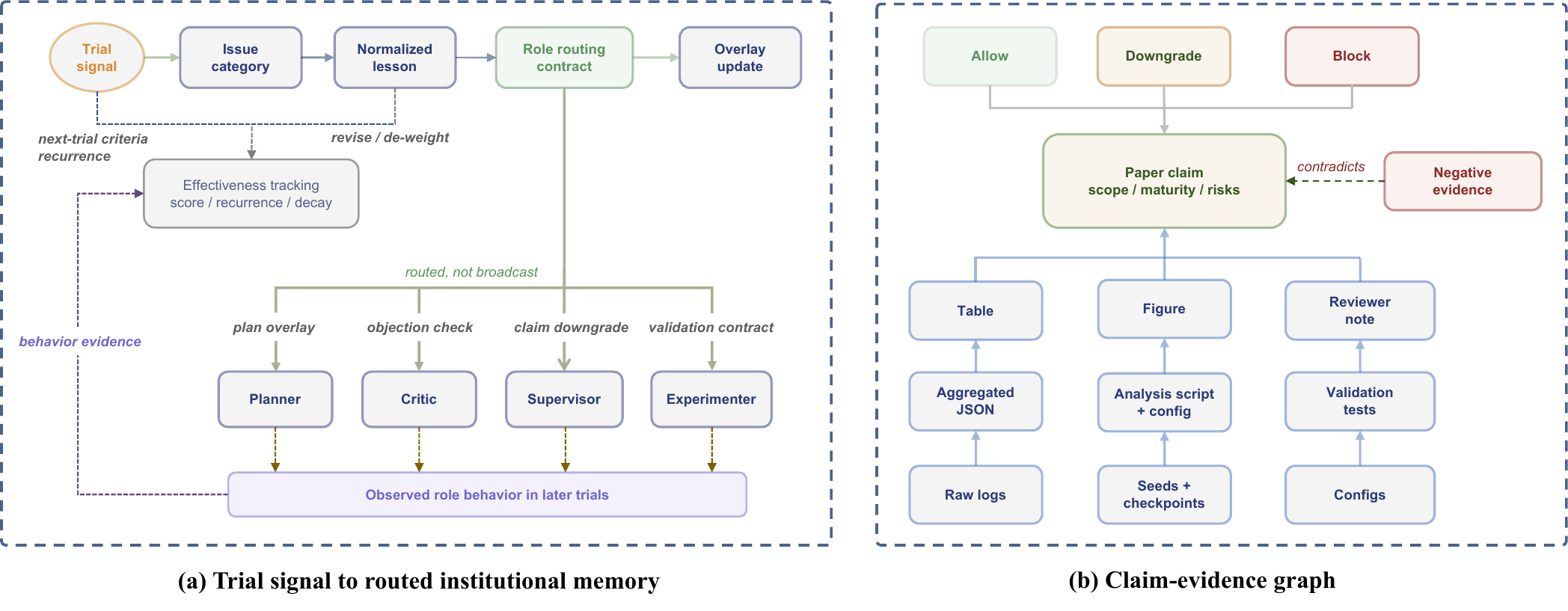}
  \caption{Memory routing and claim-evidence substrates. (a) Trial signals are normalized into issue categories and lesson records, routed to specific roles via overlay updates, and observed as later plan overlays, objection checks, claim downgrades, or validation contracts. (b) The claim-evidence graph links each prose claim to artifacts (tables, figures, reviewer notes, aggregated JSON), analysis scripts and configs, validation tests, raw logs, seeds, checkpoints, and negative evidence.}
  \label{fig:appendixsubstrates}
\end{figure}

Figure~\ref{fig:appendixsubstrates} complements the maturity diagram by showing the memory routing and claim-evidence substrates that support those checks.

\section{Extended workspace case notes}
\label{sec:extended-cases}

The following case notes provide the detailed artifacts behind the process-evidence audit. Each case is written around the same structure: trial signal, harness mechanism, and behavior update.
These traces are theory-building and stress-test evidence, not held-out validation of a completed theory.

\subsection{Evolution-memory examples}

The harness-side evidence is stored in global and per-project evolution-memory records. The important property is role routing: a repeated issue is not left as a free-form reflection note, but is assigned to affected roles with severity, frequency, suggested action, and success patterns. Table~\ref{tab:evolutionexamples} gives representative records where recurring signals become harness behavior updates.

\begin{table}[ht]
  \centering
    \vspace{6pt}
  % \small
  \scriptsize
  
  \setlength{\tabcolsep}{3pt}
  \caption{Examples of trial-to-harness-behavior conversion in evolution memory.}
  \label{tab:evolutionexamples}
  \begin{tabularx}{0.95\linewidth}{p{0.24\linewidth}p{0.4\linewidth}Y}
    \toprule
    Evolution-memory record & Recurring signal & Harness behavior update \\
    \midrule
    Supervisory lesson record &
    Post-hoc power analysis appears 5 times with high severity; conditional mutual information correlations reverse sign across dimensions; the paper's mean value 0.687 conflicts with the source value 0.6492, and Mann--Whitney $U=41.0$ conflicts with source $U=28.0$. &
    Supervisor prompts inherit checks against post-hoc justification, cherry-picking, dimension instability, and source-data mismatch. \\
    Planning lesson record &
    Random sparse-autoencoder baselines, sparsity mismatches, tautological co-occurrence measurements, and unexecuted validation experiments recur across absorption cases; one inhibition graph has precision@20=0.0 across 520 predictions, a sparsity level is unmatched in a baseline comparison, and 0/12 tests survive Bonferroni or BH-FDR correction. &
    Planner-side lessons emphasize matched controls, baseline comparisons, ablations, and execution of gatekeeper experiments before paper writing. \\
    Critique lesson record &
    Contradictory empirical results and causal language in observational analyses recur, including original positive steering correlations later contradicted by controlled matched designs. &
    Critic-side lessons route contradiction checks, causal-language audits, correction for multiple comparisons, and demand for reconciled methodology changes. \\
    Global lesson record &
    Success patterns are recorded together with failures: honest negative results recur for six iterations, stable infrastructure completes repeated task batches with zero experiment failures, and specific numbers are consistently preferred. The same memory stores probe-quality confounds such as absorption rate versus probe accuracy $\rho=-0.67$, $p<0.001$. &
    The harness preserves positive operating habits as prompt overlays rather than only accumulating warnings. \\
    \bottomrule
  \end{tabularx}
\end{table}

\subsection{Complete project traces}

The traces below are organized temporally because the claim is temporal: a signal in one iteration must change a later action. Table~\ref{tab:evolutionexamples} above gives cross-project memory evidence. The following complete traces show how a reader should inspect one project end to end. Each trace follows the same chain: project goal, iteration state, trial signal, harness decision, next behavior, and claim effect.

\subsubsection{Complete trace I: Sparse-autoencoder absorption, from writing stagnation to validation-first research}

\textbf{Project goal.} The workspace studies sparse-autoencoder feature absorption: when a parent feature absorbs child-feature behavior, how that effect should be measured, and which claims survive across layers, domains, and interventions. This is the best long-horizon case because it contains 11 iteration directories and full role artifacts across planning, experiments, supervision, writing, and reflection.

\textbf{Iteration map.}
\begin{enumerate}[leftmargin=*, itemsep=2pt]
  \item \textbf{Iterations 1--5: initial measurement and narrative formation.} The system builds an initial absorption story, writes drafts, runs targeted probes, and accumulates reflection artifacts. These iterations establish the central research object but also start the pattern that later becomes important: the paper can become smoother while source-to-paper numeric consistency remains fragile.
  \item \textbf{Iterations 6--8: writing stagnation and missing validation.} The quality trajectory stalls around 6.5. Reflection repeatedly asks for a source-to-paper validation script, but the recommendation remains a lesson rather than a hard writing gate. Iteration 8 records the ninth recommendation of the script and finds a fabricated 12.3\% hedging number where raw data gives 0.0\%. The action plan turns this into Gate 0: a 1.5-hour, zero-GPU source-to-paper cross-check must run before further writing.
  \item \textbf{Iteration 9: experiment-first break from polishing.} The project breaks the writing-only loop by executing new empirical checks: activation patching, tightened hedging analysis, conditional-mutual-information replication, and threshold sensitivity. The score rises from 6.5 to 7.0 because the system has produced new evidence rather than only a cleaner narrative.
  \item \textbf{Iteration 10: scientific progress plus evidence-boundary failure.} The iteration produces a strong probe-degradation result ($R^2=0.777$, $\rho=-1.0$, $p=0.009$), decoder-magnitude evidence (6.16 nats for first-letter and 3.98 nats for city-continent), and rate-distortion rejection across 131 pairs. The score still regresses from 7.0 to 6.5 because the paper imports new integrity errors: Table 3 confidence-interval inversion, three incompatible first-letter rates, stale 4.1$\times$ headline language, layer-multiplier mismatch, and an unverified patching sign reversal.
  \item \textbf{Iteration 11: data integrity becomes the iteration objective.} The next plan explicitly makes the iteration about data integrity and verification. The source-to-paper validation script is implemented, 51/53 checks pass, CI inversions are fixed, per-token aggregation becomes canonical, the headline changes from 4.1$\times$ to 2.7$\times$, and the 21.6\%/27.1\%/34.5\% first-letter rates are traced to distinct experimental conditions. Probe degradation becomes contribution \#1, and a 20-entity city-continent spot-check confirms 62.7\% mean recovery versus 61.9\% expected ($d=2.04$, $p<0.001$).
\end{enumerate}

\textbf{What the trace demonstrates.} The sparse-autoencoder absorption case shows the full agent-harness loop. Trial signals first change the agent's research behavior: writing gives way to experiments, then experiments give way to validation-first planning. They also change the harness boundary: source-to-paper numeric validation becomes a gate rather than an optional reflection note. The final contribution hierarchy is not the one the early drafts wanted; it is the one that survived repeated trial, critique, and validation failure mode.

\subsubsection{Complete trace II: Dynamic weight decay, from unstable control law to scoped advancement}

\textbf{Project goal.} The workspace studies dynamic weight decay: automatically changing the regularization strength during neural-network training. It is the cleanest positive trace because a concrete algorithmic defect becomes a repaired method, explicit tests, a mutated task plan, and scoped advancement.

\textbf{Iteration map.}
\begin{enumerate}[leftmargin=*, itemsep=2pt]
  \item \textbf{Iterations 0--7: idea formation, early pilots, and repeated evidence gaps.} The project builds a dynamic weight-decay story and accumulates experiments across small and medium settings. Quality moves upward and downward rather than monotonically: the quality log includes 5.5, 7.0, 5.0, 6.5, 6.75, 7.0, and later 6.5. This volatility is useful because it exposes the harness's claim-boundary function: better prose or a promising pilot does not erase unresolved controls.
  \item \textbf{Iterations 8--12: recurring control and generalization failure mode.} Reflection and evolution records keep surfacing missing ImageNet evidence, equivalence-test weakness, budget confounds, and control reliability. The evolution outcome marks missing ImageNet evidence as recurring for 7+ iterations, records equivalence tests passing only 6/12 comparisons, and proposes a preliminary smoke test before a 9-run ImageNet-100 replication plan.
  \item \textbf{Iteration 13: refinement becomes unavoidable.} Reflection records raw-log mismatches, hidden negative auxiliary-baseline results, corrupted controls, higher-regularization control gaps, and a 90-epoch ImageNet need. The important system behavior is that these signals do not get absorbed as prose caveats. They become blockers for broad advancement.
  \item \textbf{Iteration 14: repair, validation, and scoped advancement.} The supervisor path introduces a repaired controller with floor clipping, moving-average smoothing, and epoch-budget assertions. The fix passes 9/9 stability tests; the single-parameter controller budget changes from 0.0 to 90.61; ImageNet control-signal informativeness reaches 0.987; one hypothesis is no longer supported; and the ImageNet budget confound remains explicitly recorded. The plan mutates into a 14-task refinement and full-experiment program including controller repair, diagnostic CIFAR-10, unification fitting, CIFAR-100 ablations, batch-size sweeps, temporal-gate tests, alignment informativeness, ImageNet main runs, and budget-matched ImageNet controls.
\end{enumerate}

\textbf{What the trace demonstrates.} The dynamic weight-decay case shows trial-to-behavior conversion in its most direct form. A failed or unstable trial changes the algorithm, then the task plan, then the validation order, then the claim boundary. The final advancement decision is stronger because it is scoped: one hypothesis is narrowed to a three-tier taxonomy, another is treated as unsupported or uncertain, budget confounds remain visible, and the full-run plan inherits the repair obligations.

\subsubsection{Complete trace III: Diffusion-language-model acceleration, from speedup narrative to interference taxonomy}

\textbf{Project goal.} The workspace studies acceleration methods for diffusion language models and asks whether methods compose multiplicatively. It is shorter than the previous two projects, but it is the sharpest negative-evidence trace: the harness converts unsupported speedup claims into metric repair, full-scale gates, and a new thesis about interference.

\textbf{Iteration map.}
\begin{enumerate}[leftmargin=*, itemsep=2pt]
  \item \textbf{Iteration 1: paper and metric errors become experiment requirements.} Reflection fixes fabricated Wilcoxon claims, a tau$=0.0$ paradox, failure-atlas number mismatches, a quality-adjusted-speed formula inconsistency, a 6-pair overclaim where only 3 pairs are feasible, novelty overclaiming, and a speed-report mismatch for one proposed accelerator. The same action plan finds a new accept-rate error: the draft claims $\alpha=0.52$, while raw results report average accept rate 0.881 on GSM8K and 0.830 combined. Pairwise-composition evidence is only a 2-seed, 200/1319-sample pilot with per-seed range [1.292, 1.478], so the plan calls for full benchmark replication on 1319 GSM8K plus 500 MATH500 with 3 seeds and bootstrap confidence intervals.
  \item \textbf{Iteration 2: the scientific story flips from multiplication to interference.} Result debate reports 15 experiment groups, one proposed accelerator as a functional no-op around 1.16$\times$, destructive interference between two accelerators, partial interference between another accelerator pair, and an autoregressive baseline comparison where Qwen2.5-7B reaches 96\% GSM8K at 70.9--471.1 tokens per second. The old claim that speedups compose multiplicatively no longer matches the evidence.
  \item \textbf{Iteration 3: pilot/full maturity rules become explicit.} Later lessons encode the rule the earlier paper lacked: pairwise-composition evidence with $N<500$ must be labeled a pilot estimate with bootstrap intervals; core pairwise claims require full-scale $N\ge1319$ and 3 seeds. The system also separates per-token speed from output-length effects and standardizes baseline throughput before additional comparisons.
\end{enumerate}

\textbf{What the trace demonstrates.} The diffusion-language-model acceleration case shows that negative results are not dead ends. They are belief-calibration events. The harness forces unsupported statistics out of the paper, splits metric definitions, blocks overbroad pairwise claims, and changes the research question from ``how do we multiply speedups?'' to ``which mechanisms interfere, and under what evidence maturity?''

\subsubsection{Boundary trace: failed sparse-autoencoder replication, when writing quality rises as evidence collapses}

The failed sparse-autoencoder replication is not used as a clean success story. Its value is a boundary failure that exposes why data-validation gates must sit before narrative generation. Iteration 1 marks the first-letter proxy as degenerate: 26/27 GPT-2 checkpoints in one experiment and 9/10 in another return exactly 0.0, and the task-agnostic metric is negatively correlated with the first-letter benchmark ($r=-0.592$, $p=0.12$). The pilot quality is marked as not ready to proceed, and the action plan marks the pilot/full escalation error as requiring a system change: if any pilot rating is not ready to proceed, the next stage must be metric or code repair rather than scale-up.

Later, the project shows the danger of writing improvement without evidence improvement. A full component run covers 7 variants by 5 replicates, but reflection finds that writing reaches 8/10 while supervisor and critic scores fall to 4.5/10 and 5/10. The evidence base contains byte-identical replicates across nine metrics, a run manifest with 1,024 features while the paper claims 16,384 features, a missing sparsity-matched ablation, 81.6\% inactive sparse features, negative explained variance, and a canonical summary with only 3/7 variants. This trace is the cleanest warning that polished writing can move in the opposite direction from scientific maturity.

\subsubsection{Reversal trace: sparse-autoencoder hypothesis reversal, from falsified hypothesis to new framing}

The sparse-autoencoder hypothesis-reversal case shows how the system handles a result that contradicts the original story. Iteration 1 reverses one hypothesis: high-absorption features are more steerable, not less, with reported $r=+0.3548$ and $p=2.92e{-}04$. The paper reframes around ``Absorption as Steering Signature'' instead of discarding the result as a failure. Later validation complicates the new story: a controlled matched design gives $p=0.299$, one reversal is uncorrected at $p=0.015$, a steering metric saturates at layer 8, and later steering-protocol summaries report all effects at 0.0. The valuable behavior is the sequence: falsification becomes a new framing, and later validation is still allowed to downgrade that framing and create pivot signal.

\subsubsection{Pilot/full boundary trace: image-augmentation case}

The image-augmentation workspace runs a CIFAR-10/ResNet18 small pilot on a 5k subset for 10 epochs and observes a 2.68 percentage-point spread, but reflection assigns only 4.0/10 because the full-scale transition is blocked. The experiment state records 20 tracked tasks, 18 completed and 2 failed. The useful signal is not the augmentation result itself; it is the separation between task completion, pilot direction, and claim-ready evidence. A second small case, the generalization pilot, reaches a go decision with 0.88 confidence and 48.8s training while the workspace status records that execution was stopped: the experiment state tracks 8 tasks, 6 completed and 2 still running, and later synthesis artifacts still carry a pilot-mode label rather than a full 200-epoch, 7-seed, 27-combination result. Together these cases show that compute state and evidence maturity are inseparable.

\subsection{Diagnostic workspace observations}

The diagnostic workspaces are stress tests for update paths rather than clean component ablations. A no-debate setting has a clean configuration knob and reaches a pilot go decision while exposing metric sensitivity. A memory-positive setting is best read as an evolution-follow-up case rather than a clean memory ablation: a recurring activation-patching lesson is later answered by a pilot with 9/9 patching checks and 67.3\% mean recovery. A memory-negative setting is strongest as a boundary case because the quality gate hard-blocks a missing review score and rolls the workspace back. A validation-removal setting is config-only in this audit, and a no-revision setting is a stagnation/escalation case rather than a clean revision-off component test.

\textbf{Details used in the main audit.} The no-debate setting ran a synthetic absorption pilot to a go decision; the trained sparse autoencoder showed higher absorption than the random baseline under the overlap method (0.50 versus 0.25), while the ablation method gave 1.00 for both, exposing a measurement-method sensitivity that debate should expose earlier. The memory-positive setting records explicit validation against evolution lessons: activation patching succeeds on 9/9 checks with 67.3\% mean recovery, but later reflection still records multiple-comparison, baseline, figure, and circularity issues. The memory-negative setting contains a runtime boundary artifact: the quality gate hard-blocks iteration 5 because it has no review score and rolls the system back to review.

\textbf{System-evolution details.} The evolution-memory records normalize reflection outputs into issue categories, severities, affected roles, suggestions, statuses, and success patterns. In the audited workspace set, 12 of 13 workspaces have non-empty outcome records, and the central ledger contains 173 outcome records (the issue-pattern category mix is reported in Section~\ref{sec:process-evidence} and not repeated here). Representative process failures include missing rendered figures, placeholder references, paper-length overflow, absent code-release plans, underpowered experiments, stale artifacts, paper/LaTeX desynchronization, incomplete GPU telemetry with empty timing fields, non-self-contained evidence bundles with absolute paths, and external synchronization failures. Those failures become routed lesson records for planning, experiment-running, supervision, critique, skepticism, writing, and editing roles. Table~\ref{tab:appendixablations} summarizes how the diagnostic workspace settings are used in this audit.

\begin{table}[ht]
  \centering
    \vspace{6pt}
  \small
  \setlength{\tabcolsep}{3pt}
  \caption{Appendix diagnostic-workspace status.}
  \label{tab:appendixablations}
  \begin{tabularx}{0.95\linewidth}{p{0.20\linewidth}p{0.25\linewidth}Y}
    \toprule
    Setting & Use in this draft & Trace value \\
    \midrule
    No-debate setting & Perspective diagnostic. & Clean debate-disabled setting plus metric-sensitivity trace: 0.50 vs. 0.25 under one metric, 1.00 vs. 1.00 under another. \\
    Memory-positive setting & Evolution-follow-up case. & Recurring activation-patching lesson becomes a validation block with 9/9 checks and 67.3\% mean recovery. \\
    Validation-removal setting & Validation-removal setting. & Config-only stress-test setting in this audit; useful for protocol design, not outcome evidence. \\
    No-revision setting & Stagnation/escalation case. & Records score stagnation, recurring issues, topic drift, and action plans that were not enforced. \\
    Memory-negative or weakened settings & Boundary and memory diagnostics. & Hard quality-gate rollback and repeated proxy/control failures expose what must be routed into gates and planner prerequisites. \\
    \bottomrule
  \end{tabularx}
    \vspace{1pt}
\end{table}

\subsection{Recovered-failure registry}
\label{sec:appendix-recovered-failures}

Table~\ref{tab:appendix-recoveredfailures} lists the natural failure classes used for the recovered-failure audit referenced in Section~\ref{sec:process-evidence}. These are existing failures that the harness blocked, downgraded, or routed into repair, not newly injected tests.

\begin{table}[ht]
  \centering
    \vspace{6pt}
  \small
  \setlength{\tabcolsep}{2pt}
  \caption{Recovered failures from natural workspace traces. These are existing failures that the harness blocked, downgraded, or routed into repair, not newly injected tests.}
  \label{tab:appendix-recoveredfailures}
  \resizebox{0.95\linewidth}{!}{%
  \begin{tabularx}{\linewidth}{p{0.14\linewidth}p{0.22\linewidth}p{0.24\linewidth}p{0.20\linewidth}Y}
    \toprule
    Failure class & Signal & Audit artifact type & Harness catch & Later update \\
    \midrule
    Duplicate result files & 4 of 5 component replicates byte-identical to another method across nine metrics. & Reflection note from failed sparse-autoencoder replication & Supervisor and critic downgraded evidence despite high writing score. & Duplicate detection and single-source analysis prerequisites. \\
    CI inversion & 5 of 7 confidence intervals did not contain point estimates. & Reflection note from sparse-autoencoder absorption case & Reflection marked the error critical and tied it to missing validation. & Source-to-paper validation script; CI fixes in the next iteration. \\
    Stale headline number & 4.1\(\times\) ratio persisted after data changed. & Reflection note and revised experiment section from sparse-autoencoder absorption case & Source-to-paper validation exposed stale prose. & Headline reduced to 2.7\(\times\) under quality-gated aggregation. \\
    Feature-count mismatch & Run used 1024 features while paper claimed 16{,}384. & Reflection note from failed sparse-autoencoder replication & Feature-count verification became a prerequisite. & Claim generation blocked until manifest/run alignment. \\
    Unsupported statistics & Fabricated Wilcoxon values and $\alpha=0.52$ vs. measured accept rate 0.881. & Action plan and result-debate record from diffusion-language-model acceleration case & Reflection and result debate rejected unsupported statistics. & Paper reframed around interference and full-scale gates. \\
    \bottomrule
  \end{tabularx}}
\end{table}

\section{Review-artifact and transition statistics}
\label{sec:appendix-review-transition-stats}

The generated-review archive contains reviewer-like artifacts for workspace paper drafts. We use them as process pressure tests only. They are not human peer-review decisions and do not validate the domain claims of the drafts. Their value is to expose whether external objections would become claim-boundary or validation work. We also align these generated reviews with available internal \sibyl{} supervisor reviews when a structured review record exists for the same project iteration. Table~\ref{tab:reviewstats} separates score calibration from transition structure because neither score should be optimized directly.

\begin{table}[ht]
  \centering
    \vspace{6pt}
  % \footnotesize
  \small
  \setlength{\tabcolsep}{3pt}
  \caption{Review-artifact parsing summary. Numeric review scores are treated as calibration diagnostics and audit signals, not as scientific-quality targets.}
  \label{tab:reviewstats}
  \resizebox{\linewidth}{!}{%
  \begin{tabularx}{\linewidth}{p{0.23\linewidth}p{0.24\linewidth}p{0.24\linewidth}Y}
    \toprule
    Statistic & Conservative reviewer & Rubric reviewer & Long-form reviewer \\
    \midrule
    Artifact count & 51 generated reviews over the same project-iteration set. & 51 generated reviews plus 51 metadata JSON files, all marked completed. & 51 MHTML reviews plus 51 PDFs; MHTML contains the review text. \\
    Coverage & 11 workspaces and 51 project-iteration snapshots. & 11 workspaces and 51 project-iteration snapshots. & 11 workspaces and 51 project-iteration snapshots. \\
    Native score schema & Complete numeric reviews for 35/51 snapshots; desk-only outputs for 16/51. & Rubric subscores for 51/51 snapshots; numeric overall for 18/51. & Long-form assessment with no native numeric score; automated text parsing assigns coarse ordinal labels to 31/51 assessments and leaves 20/51 unclear. \\
    Overall-score distribution & 31/35 overall scores are 3; 4/35 are 2; 0/35 are 4 or higher. & 12/18 numeric overall scores are 4 or higher; mean 3.94. & Text-inferred labels: 10 reject, 4 major revision, 14 borderline/workshop, 3 positive, 20 unclear; mapped to scores 2/3/4 only for Figure~\ref{fig:reviewprocessdiag}. \\
    Parsed subscore calibration & Mean 2.31 across 140 parsed dimension scores. & Mean 2.65 across 204 parsed dimension scores; paired subscores average 0.34 points higher than the conservative reviewer. & No native subscore schema; inferred ordinal mean is 2.65 over the 31 classified text assessments. \\
    Adjacent scored transitions & 20/22 consecutive overall-score transitions are flat; 1 rises and 1 falls. & Only 3 consecutive numeric-overall pairs are available: 1 flat, 1 rise, 1 fall. & Inferred ordinal transitions are non-monotone across 14 adjacent classified pairs: 3 flat, 5 rise, 6 fall. \\
    Interpretation & Conservative gate; useful for desk-reject and evidence-boundary signals. & More positive calibration; useful as a different review surface, not as a replacement for artifact audit. & Useful for detailed qualitative objections; inferred ordinal trend claims require caution. \\
    \bottomrule
  \end{tabularx}}
\end{table}

\begin{figure}[ht]
  \centering
  \vspace{4pt}
  \includegraphics[width=\linewidth]{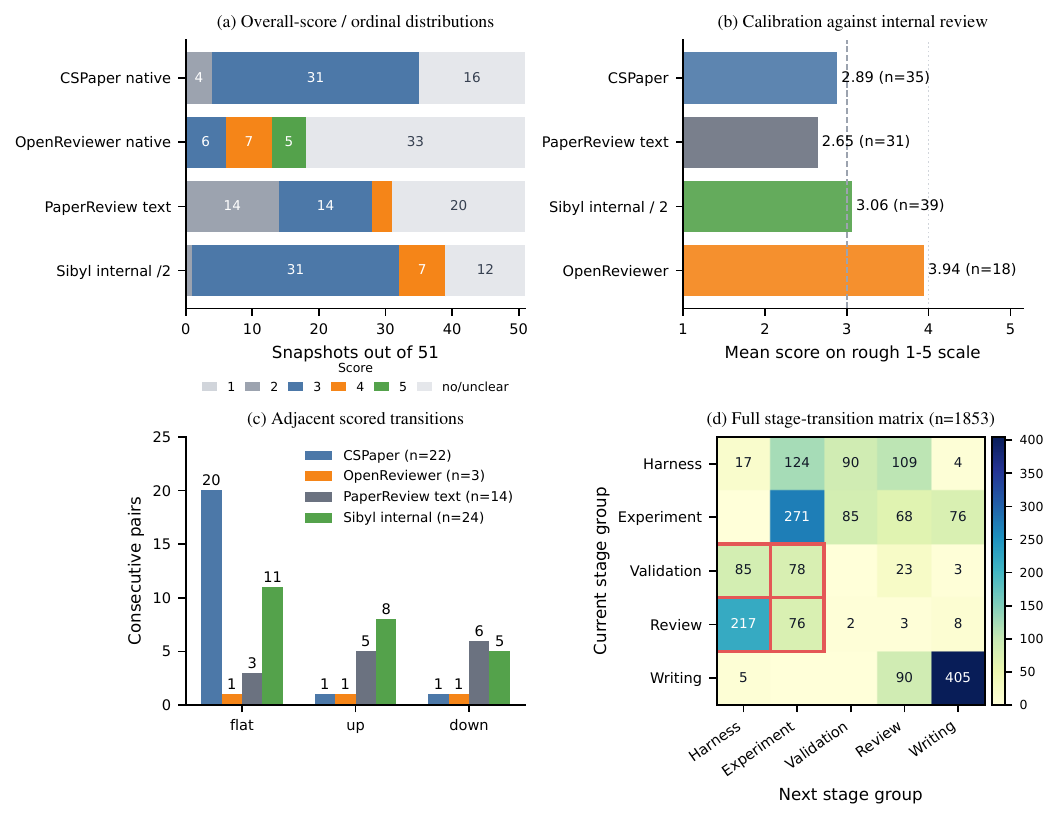}
  \caption{Review-artifact calibration and stage-transition counts (appendix view of the data discussed in Section~\ref{sec:process-evidence}). (A) Two native 1--5 generated-review scores, long-form-reviewer text-inferred ordinal labels, and internal \sibyl{} supervisor scores divided by two and rounded. Gray no/unclear regions mark missing numeric overall scores or unclassified text assessments. (B) Internal \sibyl{} scores sit closer to the conservative reviewer than to the rubric reviewer; the long-form reviewer is text-inferred only. (C) Adjacent score transitions differ by review surface; no single review score should be optimized directly. (D) The full stage-transition matrix shows writing self-loops, and review or validation stages routing work back to harness and experiment stages.}
  \label{fig:reviewprocessdiag}
\end{figure}

Internal supervisor reviews are not directly comparable to generated reviewer outputs because they use a 10-point process-review scale and often include role-specific evidence gates. They are still useful as a calibration reference. Among the 51 generated-review snapshots, 39 had a score-bearing \sibyl{} supervisor review; the mean internal score was 6.13/10, with available dimension-score means of 6.74/10 for novelty, 5.38/10 for experiments, 5.41/10 for soundness, and 5.66/10 for reproducibility across 34 reviews. For the Figure~\ref{fig:reviewprocessdiag} distribution panel, we divide these scores by two and round them, giving 1 snapshot at score 2, 31 at score 3, and 7 at score 4. Using the rough conversion internal/2 to compare against 5-point generated scores, the paired difference internal/2 minus the conservative reviewer averaged +0.14 over 27 pairs, while internal/2 minus the rubric reviewer averaged -0.83 over 12 pairs. This supports the calibration claim only; it should not be read as a unified quality metric.

For Figure~\ref{fig:scorebehaviordiag}, we separately parse internal reflection-derived action plans and the next iteration's task plans. Table~\ref{tab:scoreactionstats} reports the parsed action-plan rows and visible next-plan task mix. Recommendation categories are heuristic and multi-label because one focus item can ask for both a new experiment and a paper change; next-plan task categories use one primary category per task. The statistic is therefore a process diagnostic rather than an exact causal effect estimate.

\begin{table}[ht]
  \centering
    \vspace{6pt}
  % \footnotesize
  \small
  
  \setlength{\tabcolsep}{2pt}
  \caption{Internal review-to-action parsing summary across 37 parseable action-plan rows in the 12 audited traces. Score movement uses a \(\pm 0.25\)-point threshold on the 10-point internal review scale when a prior score exists.}
  \label{tab:scoreactionstats}
  \resizebox{0.95\linewidth}{!}{%
  \begin{tabularx}{\linewidth}{p{0.16\linewidth}p{0.08\linewidth}p{0.10\linewidth}p{0.14\linewidth}p{0.14\linewidth}Y}
    \toprule
    Score movement & Rows & Mean delta & High-severity issues/row & Focus items/row & Visible next-plan task mix \\
    \midrule
    Down & 10 & -0.82 & 8.7 & 8.4 & 88 tasks in 7 visible next plans: 56 experiment/control, 17 validation/artifact, 13 harness/system, 1 claim/writing, 1 other. \\
    Flat & 6 & 0.00 & 6.5 & 5.8 & 48 tasks in 5 visible next plans: 28 experiment/control, 12 validation/artifact, 7 harness/system, 1 claim/writing. \\
    Up & 10 & +0.70 & 4.0 & 7.5 & 55 tasks in 5 visible next plans: 31 experiment/control, 13 validation/artifact, 9 harness/system, 1 claim/writing, 1 other. \\
    No prior score & 11 & -- & 3.8 & 5.9 & 79 tasks in 8 visible next plans: 57 experiment/control, 10 validation/artifact, 10 harness/system, 1 claim/writing, 1 other. \\
    \bottomrule
  \end{tabularx}}
    \vspace{6pt}
\end{table}

The artifact coverage itself is an audit signal. The review set contains 51 reviewed project-iteration pairs, while the local collected-paper tree used for this paper contains 46 project-iteration artifact folders. Nine reviewed iterations are absent from that local collection and four local collected iterations are not reviewed. In the image-augmentation case, the latest collected Markdown draft makes full-scale 200-epoch claims, while the reviewed source still describes pilot 10-epoch, 100-sample, single-seed results. This is not evidence about reviewer accuracy; it is evidence that a research harness needs artifact synchronization checks before interpreting any review score. Table~\ref{tab:stagetransitions} reports the raw stage-transition counts used to interpret the process shape.

\begin{table}[ht]
  \centering
  % \footnotesize
  \small
  \setlength{\tabcolsep}{2pt}
  \caption{Raw stage-transition counts in workspace event logs. Counts are from 1{,}853 stage-end records across 12 audited traces and may include resume/checkpoint repetition.}
  \label{tab:stagetransitions}
  \begin{tabularx}{0.95\linewidth}{p{0.25\linewidth}p{0.10\linewidth}Y}
    \toprule
    Transition group & Count & Interpretation \\
    \midrule
    Writing \(\rightarrow\) writing & 405 & Writing and revision can absorb many loops without necessarily improving evidence maturity. \\
    Experiment \(\rightarrow\) experiment & 271 & Pilot/full runs, retries, and parallel experiment batches dominate execution volume. \\
    Review \(\rightarrow\) harness & 217 & Review and debate frequently return control to planning, reflection, or other harness stages. \\
    Harness \(\rightarrow\) experiment & 124 & Planning and reflection often route lessons back into experimental work. \\
    Harness \(\rightarrow\) review & 109 & Idea and result debates are frequent harness-side review surfaces. \\
    Writing \(\rightarrow\) review & 90 & Draft production commonly triggers final review or quality assessment. \\
    Harness \(\rightarrow\) validation & 90 & Planning/reflection can route work into explicit decision or quality gates. \\
    Experiment \(\rightarrow\) validation & 85 & Trial outputs are often checked by validation or decision stages. \\
    Validation \(\rightarrow\) harness & 85 & Validation gates can return immature evidence to planning, reflection, or other harness stages. \\
    Validation \(\rightarrow\) experiment & 78 & Gates can force additional experiments instead of allowing narrative advancement. \\
    Review \(\rightarrow\) experiment & 76 & Review/debate objections often require new or repaired experiments. \\
    \bottomrule
  \end{tabularx}
\end{table}

\section{System comparison details}
\label{sec:appendix-systemcomparison}

\begin{figure}[ht]
  \centering
  \vspace{6pt}
  \includegraphics[width=\linewidth,height=0.18\textheight,keepaspectratio]{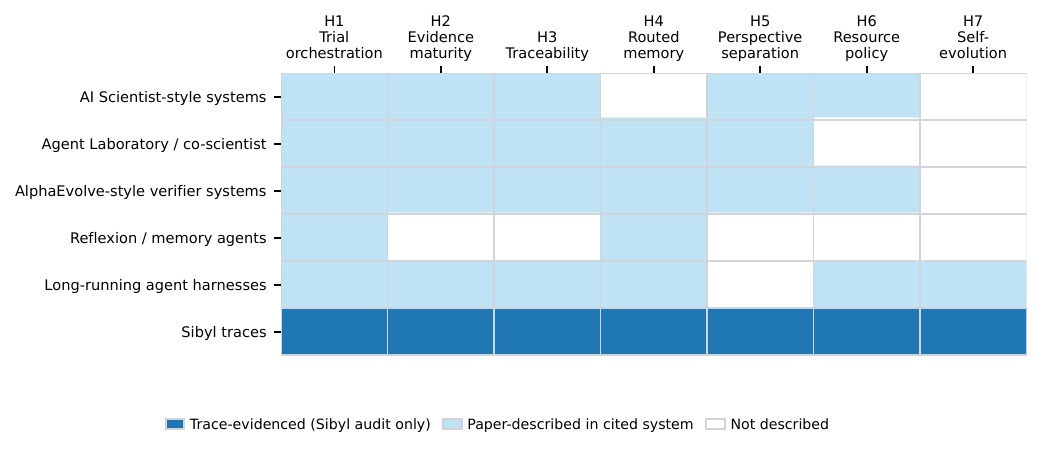}
  \caption{Documented support for the H1--H7 commitments under two non-comparable evidence levels. \textit{Trace-evidenced} cells (\sibyl{} only) come from the audited workspace artifacts in this paper. \textit{Paper-described} cells come from cited system descriptions and indicate that the published text exposes the ingredient, not that the trace-level update path was audited. \textit{Not described} indicates the cited description does not expose the ingredient.}
  \label{fig:systemcomparison}
  \vspace{1pt}
  
\end{figure}

Figure~\ref{fig:systemcomparison} and Table~\ref{tab:appendix-systemcomparison} compare what different lines of work make observable under the H1--H7 commitments. The comparison is not a leaderboard. It separates two distinct evidence levels on purpose: prior systems are scored from their published descriptions (\textit{paper-described}: the system text exposes the relevant ingredient) versus our own \sibyl{} audit (\textit{trace-evidenced}: the workspace artifacts contain a recoverable signal-to-update path). The two levels are not directly comparable, and a \textit{paper-described} entry should not be read as weaker engineering. The point is that auditing trace-level update paths requires access to internal artifacts that we have only for \sibyl{}.

\begin{table}[ht]
  \centering
  \vspace{2pt}
  \small
  \caption{Table form of Figure~\ref{fig:systemcomparison}. \textit{TE} = trace-evidenced (audited workspace artifacts; available only for \sibyl{}). \textit{PD} = paper-described (the cited system description exposes the ingredient). \textit{ND} = not described. The two levels are not directly comparable; \sibyl{}'s entries reflect what the audit could recover from preserved traces, not a comparative performance claim.}
  \label{tab:appendix-systemcomparison}
  \begin{tabular}{lccccccc}
    \toprule
    System & H1 & H2 & H3 & H4 & H5 & H6 & H7 \\
    \midrule
    AI Scientist-style systems & PD & PD & PD & ND & PD & PD & ND \\
    Agent Laboratory / co-scientist & PD & PD & PD & PD & PD & ND & ND \\
    AlphaEvolve-style verifier systems & PD & PD & PD & PD & PD & PD & ND \\
    Reflexion / memory agents & PD & ND & ND & PD & ND & ND & ND \\
    Long-running agent harnesses & PD & PD & PD & PD & ND & PD & PD \\
    \sibyl{} traces (this paper) & TE & TE & TE & TE & TE & TE & TE \\
    \bottomrule
  \end{tabular}
  \vspace{2pt}
  
\end{table}

\section{Evaluation protocols}
\label{sec:appendix-eval-protocols}

These six protocols are how the framework would be tested at scale. Only the first protocol is exercised in this paper; the remaining five are the evaluation contract this framework invites future work to fulfill, with held-out harnesses, independent annotators, prospective injected failures, and public artifact bundles.

\paragraph{Retrospective trial-to-behavior audit (used in this paper).} Given a completed workspace, reconstruct the project goal, trial signals, harness mechanisms triggered, behavior updates, maturity labels, negative evidence, and evidence paths. The audit fails if the claimed behavior update cannot be tied to artifacts.

\paragraph{Prospective fixed-budget study.} Give several harness designs the same research question, compute budget, and human review budget. Compare maturity gain, trace fidelity, unsupported claims, negative-result handling, compute reliability, and human audit burden.

\paragraph{Injected-failure stress test.} Insert controlled failures such as duplicate result files, missing outputs, stale tables, inconsistent feature counts, unsupported statistics, or pilot/full mismatches. A strong harness should block or downgrade narrative generation and point to the offending artifact.

\paragraph{Cross-project memory test.} Let one project expose a failure mode, then start another project where the same failure could recur. Measure whether the lesson is retrieved, routed to the right role, and converted into a changed plan or validation check.

\paragraph{Harness-evolution test.} Let one project expose a process failure such as missing telemetry, stale figure assets, absolute artifact paths, or paper/LaTeX desynchronization. Start later projects where the same failure could recur. Measure whether the harness adds a gate, sentinel, repair task, artifact contract, or scheduler policy, and whether recurrence rate falls.

\paragraph{Perspective and efficiency ablations.} Remove or weaken skeptic, methodologist, supervisor, validation, memory, or scheduling components under controlled budgets. The outcome should not be only task completion. It should include overclaim rate, missed validation failures, disagreement-to-action conversion, maturity gain, and audit burden.

\section{Governance details}
\label{sec:appendix-governance-details}

A research harness makes autonomous trial-and-error safer and more auditable while preserving human responsibility. If optimized poorly, it becomes a more efficient machine for plausible weak science. The main risks are scientific spam, rhetorical overclaiming, data integrity failures, metric gaming, unsafe trial-and-error in high-risk domains, negative memory suppressing valid exploration, perspective theater, and self-evolution drift.

The corresponding mitigations are claim-evidence gates, AI-use disclosure, human accountability, negative-evidence reporting, domain-specific safety gates, hidden injected failures, protected integrity constraints, and periodic audit of memory overlays. Table~\ref{tab:governance} summarizes the risk-mitigation pairs. The most important policy point is simple: human authors must be able to understand, defend, and rewrite the final submission. Autonomous systems can help create evidence traces and drafts; they should not be treated as accountable authors.

\begin{table}[ht]
  \centering
  \vspace{1pt}
  \small
  \setlength{\tabcolsep}{3pt}
  \renewcommand{\arraystretch}{1.1}
  \caption{Governance risks and mitigations. Each row reads as: when the harness creates the failure mode in column~2, the mitigation in column~3 keeps it auditable.}
  \label{tab:governance}
  \begin{tabularx}{0.95\linewidth}{p{0.24\linewidth}p{0.32\linewidth}Y}
    \toprule
    Risk & When it arises & Mitigation \\
    \midrule
    Scientific spam & Paper completion is treated as success. & Claim-evidence gates and validation-first writing. \\
    Rhetorical overclaiming & Writer holds full claim authority. & Maturity-labeled claim registry consumed by writers. \\
    Metric gaming & A single harness score becomes the proxy. & Multi-dimensional labels and hidden injected failures. \\
    Negative-memory suppression & Lessons are over-generalized into bans. & Time decay, source links, and re-opening criteria. \\
    Self-evolution drift & Overlays or gates optimize polish over integrity. & Periodic memory audits against an integrity baseline. \\
    Unsafe self-modification & Repair tasks weaken protected constraints. & Human-reviewed protected files, regression tests, and rollback logs. \\
    \bottomrule
  \end{tabularx}
\end{table}

\end{document}